\documentclass[a4paper]{article}
\begin{document}
\begin{flushright}
\hfill CAMS/98-01\\
\hfill hep-th/9806046\\
\end{flushright}
\vspace{1cm}
\begin{center}
\baselineskip=16pt
{\Large\bf Remarks on the Spectral Action Principle}
\vskip 2cm
{\bf Ali H. Chamseddine} \\
\vskip 0.5cm
{\em Center for Advanced Mathematical Sciences\\
and\\
Physics Department\\
American University of Beirut \\
Beirut, Lebanon}\\
E-mail chams@aub.edu.lb
\vskip 0.2cm
\end{center}
\vskip 1cm
\begin{abstract}
The presence of chiral fermions in the physical Hilbert space
implies consistency conditions on the spectral action. These
conditions are equivalent to the absence of gauge and gravitational
anomalies. Suggestions for the fermionic part of the spectral
action are made based on the supersymmetrisation of the bosonic
part.
\end{abstract}
\bigskip
\bigskip
\newpage
\section{Introduction}

In noncommutative geometry \cite{Connes, Cobook} the interplay between
geometry and physics is
possible \cite{CoLo, Kastler, CFF} because a noncommutative space is defined
by a spectral triple $\left( \mathcal{A},\mathcal{H},D\right) \,$
where $\mathcal{A}$ is an algebra
of operators, $\mathcal{H}$ a Hilbert space of states, $D$ is a compact
operator acting on $\mathcal{H}$. When the space is supplied with a real
structure $J$ \cite{Co95}, an automorphism on elements of the algebra
is equivalent to
changing some initial metric $D_{0}$ to
\begin{equation}
D=D_{0}+A+JAJ^{-1}
\end{equation}
where $A=\sum_{i}a_{i}\left[ D,b_{i}\right] $ is a one form. It is
conjectured in \cite{ACAC} that the dynamics of the geometrical
fields encoded in the metric operator $D$ is given by
\begin{equation}
Tr\left( F\left( D^{2}/m^{2}\right) \right) +\left( \psi ,D\psi \right)
\label{spec}
\end{equation}
where $F$ is an arbitrary function of $D$, the trace ``$Tr$'' is
taken over the Hilbert space of states, $\psi $ is a collection of
states in the Hilbert space, and $\left( .,.\right) $ defines an
inner product in the Hilbert space. The mass scale $m$ is usually
related to a cut-off scale. When this principle is used for the
noncommutative space defined by the fermionic spectrum of the
standard model, the spectral action gives, in the low energy limit,
the standard model with all of its detailed structure \cite{ACAC}.

In the considerations of \cite {ACAC} two important points were not
dealt with. The first has to do with the fact that states in the
Hilbert space are chiral fermions. As the eigenvalue problem is
only well defined for Dirac fermions \cite{Fuji, Bala, Alv} one
must  define the trace over the chiral states only. We shall show
that such a restriction makes the trace function in (\ref{spec})
non-invariant under chiral rotation, unless certain conditions on
the fermionic representations of states in the Hilbert space are
satisfied. These conditions coincide with the cancellation of gauge
and gravitational anomalies. The second point not studied, is the
obvious asymmetry between the bosonic and fermionic parts of the
spectral action. The fermionic part is linear in $D$ while the
bosonic part involves a general function $F$ of $D$. We shall
investigate this question, to find out whether more general
fermionic actions are possible, by studying a supersymmetrization
of some of the higher order bosonic terms obtained in the heat
kernel expansion. The plan of this paper is as follows. In section
two, an expression for the anomaly of the spectral action is
derived. In section three higher order terms in the bosonic part of
the spectral action of a simple Yang-Mills system, are obtained. In
section four we supersymmetrize these higher order terms and
postulate a general form for the fermionic action. Section five
contains the conclusions.

\section{Anomalies in the spectral action}

Fermions that appear in the physical Hilbert space in any realistic
model must be chiral. The grading operator $\gamma $ is such that
\begin{eqnarray}
\gamma \psi _{\pm } &=&\pm \psi _{\pm } \nonumber\\
\gamma D\psi _{\pm } &=&\mp D\gamma \psi _{\pm }
\end{eqnarray}
which implies that $D\psi _{\pm }$ have opposite chirality to $\psi
_{\pm }$. Therefore, for chiral fermions with non-zero eigenvalues,
 one cannot set the eigenvalue problem for $D$ but only for $D^{2}$. We can write
\begin{equation}
D^{2}\psi _{n\pm }=\lambda _{n}\psi _{n\pm },\qquad \lambda _{n}=\lambda
_{n}^{*}
\end{equation}
where
\begin{equation}
\lambda _{n}=\left( \psi _{n-}\left| D^{2}\right| \psi _{n+}\right) .
\end{equation}
Assuming that all fermions in the spectrum have positive chirality
(negative chirality fermions could always be written as the
conjugate of  positive chirality fermions). The bosonic spectral
action in this case could be written as
\begin{equation}
I_{b}=Tr\left( F(D^{2})\right) =\sum\limits_{n}\left( \psi _{n-}\left|
F(D_{-}D_{+}\right| \psi _{n+}\right)
\label{bosact}
\end{equation}
where $D^{2}$ is replaced by $D_{-}D_{+}$ when acting on $\psi
_{n+}$.

Let us now consider the behavior of this action under chiral
transformations. The fermionic action is invariant under chiral rotations
\begin{eqnarray}
\left| \psi _{+}\right) &\rightarrow &e^{i\theta \gamma _{5}}\left| \psi
_{+}\right) \nonumber\\
\left( \psi _{+}\right| &\rightarrow &\left( \psi _{+}\right| e^{i\theta
\gamma _{5}}
\end{eqnarray}
which implies that
\begin{equation}
\left( \psi _{+}\left| D\right| \psi _{+}\right) \rightarrow \left( \psi
_{+}\left| e^{i\theta \gamma _{5}}De^{i\theta \gamma _{5}}\right| \psi
_{+}\right) .
\end{equation}
It is a simple matter to see that
\[
e^{i\theta \gamma _{5}}De^{i\theta \gamma _{5}}=D
\]
implying the invariance of the fermionic action.

Under chiral rotations the bosonic action (\ref{bosact}) transforms
as
\begin{eqnarray}
I_{b} &\rightarrow &Tr\left\{ \sum\limits_{n}\left( e^{-i\theta
^{a}T^{a}\gamma _{5}}\psi _{n-}\right| F(D_{-}D_{+})e^{i\theta
^{a}T^{a}\gamma _{5}}\left| \psi _{n+}\right) \right\} \nonumber\\
&=&Tr\left\{ e^{2i\theta ^{a}T^{a}\gamma _{5}}\left( \psi _{n-}\right|
F(D_{-}D_{+})\left| \psi _{n+}\right) \right\}
\label{bostra}
\end{eqnarray}
where $T^{a}$are matrix representations for the fermions. We now
use the heat kernel expansion for the function $F$ \cite{Gilkey}
\begin{equation}
F\left( P\right) =\sum\limits_{n\geq 0}f_{n}e_{n}\left( P\right)
\end{equation}
where $e_{n}\left( P\right) $ are geometric invariants whose trace
gives the Seeley-de Witt coefficients for the operator
$P=D_{-}D_{+}$. Under an infinitesimal transformation equation
(\ref{bostra}) simplifies to
\begin{equation}
I_{b}\rightarrow I_{b}+\sum\limits_{n\geq 0}f_{n}Tr\,\left( 2i\theta
^{a}T^{a}\gamma _{5}e_{n}\left( P\right) \right) .
\end{equation}
Because of the presence of $\gamma _{5}$ in the trace, the first
non-vanishing gauge term comes from $e_{4}\left( P\right) $ and is of the
form
\begin{eqnarray}
&&2i\theta ^{a}Tr\,\left( \gamma _{5}\gamma ^{\mu \nu }\gamma ^{\rho \sigma
}T^{a}T^{b}T^{c}\right) G_{\mu \nu }^{b}G_{\rho \sigma }^{c} \nonumber\\
&=&i\epsilon ^{\mu \nu \rho \sigma }\theta ^{a}G_{\mu \nu }^{b}G_{\rho
\sigma }^{c}Tr\,\left( T^{a}\left\{ T^{b},T^{c}\right\} \right) .
\end{eqnarray}
Therefore the bosonic action is non-invariant except when the condition on
the fermionic group representations
\begin{equation}
Tr\,\left( T^{a}\left\{ T^{b},T^{c}\right\} \right) =0
\end{equation}
is satisfied. We can relate the anomaly generating term to non-conservation
of currents. Let $\theta =\theta ^{a}\left( x\right) T^{a}$, then
\begin{equation}
\left( \psi _{+}\right| D\left| \psi _{+}\right) \rightarrow \left( \psi
_{+}\right| D\left| \psi _{+}\right) -i\theta ^{a}\partial _{\mu }\left(
\psi _{+}\right| \gamma ^{\mu }T^{a}\left| \psi _{+}\right)
\end{equation}
where we have used the identity
\[
e^{i\theta \gamma _{5}}De^{i\theta \gamma _{5}}=D+i\gamma ^{\mu }\gamma
_{5}\partial _{\mu }\theta
\]
and integrated by parts. Defining the fermionic current
\begin{equation}
j^{\mu a}=\left( \psi _{+}\left| \gamma ^{\mu }T^{a}\right| \psi _{+}\right)
\end{equation}
we then have
\begin{equation}
\partial _{\mu }j^{\mu a}=-\epsilon ^{\mu \nu \rho \sigma }G_{\mu \nu
}^{b}G_{\rho \sigma }^{c}Tr\,\left( T^{a}\left\{ T^{b}T^{c}\right\}
\right) .
\end{equation}

There is also one gravitational term that comes from $e_{4}\left( P\right) $
\begin{equation}
\epsilon ^{\mu \nu \rho \sigma }\theta ^{a}Tr\,\left( T^{a}\right) R_{\mu
\nu }^{cd}R_{\rho \sigma cd} .
\end{equation}
Therefore gravitational anomalies are absent if one further imposes the
condition \cite{A-W}
\begin{equation}
Tr\left( T^{a}\right) =0 .
\end{equation}

One can make similar analysis for anomalies on higher dimensional
manifolds. It can be immediately seen that in ten dimensions the first
non-vanishing trace in the heat-kernel expansion would result from a term of
the form
\begin{equation}
Tr\left( \gamma _{11}\gamma ^{\mu _{1}\mu _{2}}\ldots \gamma ^{\mu
_{9}\mu_{10}} T^{a_{1}}\cdots T^{a_{5}}\right) F_{\mu _{1}\mu
_{2}}^{a_1}\cdots F_{\mu_{9}\mu_{10}}^{a_5}
\end{equation}
which is obviously related to the well known hexagon diagram
\cite{A-W} .

We conclude this section by observing that chiral gauge
anomalies arise in the path integral formulation because
the fermionic measure is not invariant under chiral rotations
\cite{Fuji, Bala, Alv, A-W} ,
while in the spectral action they arise because of
the non-invariance of the trace operator. The anomaly cancellation
conditions are, however, identical in both cases.

\section{Fermionic part of the spectral action}

Looking at the spectral action (\ref{spec}) one cannot fail to
notice the asymmetry between the bosonic and fermionic parts. The
fermionic part has the simple form $\left( \psi \left| D\right|
\psi \right) $ while the bosonic part involves some (non-linear)
function of the Dirac operator. If one hopes for some symmetry
between the two parts of the action, there must exist a relation
between them. As an example, in supersymmetric theories the bosonic
and fermionic parts of the action are completely related to each
other. This tends to indicate that the fermionic action must be
more complicated than the simple form in (\ref{spec}). A first
guess is to take
\begin{equation}
I_{f}=\left( \psi \left| G(D)\right| \psi \right)
\end{equation}
where $G\left( D\right) $ is some (odd) function of $D$ (as follows
from chirality). Although this  is not the most general form
because one can also write terms which are not quadratic but
quartic or higher in the fermionic fields, it is the simplest form
that involves an arbitrary function. To get a concrete idea on the
possible structure of the fermionic terms we shall consider the
supersymmetrization of the bosonic part of the spectral action
associated with a simple noncommutative space. The spectral triple
we are interested in is given by
\begin{eqnarray}
\mathcal{A} &=&C^{\infty }(M)\otimes M^{N}(C) \nonumber\\
\mathcal{H} &=&L^{2}\left( M, S\right) \otimes M_N(C)\nonumber \\
D_{0} &=&\gamma ^{\mu }\partial _{\mu }\otimes 1_{N}
\label{space}
\end{eqnarray}
which defines the product of a continuous four dimensional manifold $M$
times the algebra of $N\times N$ matrices. We add a real structure $J$
(the charge conjugation operator) and supplement the fermions with their
conjugate elements. The space of metrics has a natural foliations into
equivalence classes. The internal fluctuations of a given metric are given by
\begin{equation}
D=D_{0}+A+JAJ^{-1}
\end{equation}
where $A=\sum\limits_{i}a_{i}\left[ D,b_{i}\right] $, $a_{i,}b_{i}\in
\mathcal{A}$.
Automorphisms on elements of $\mathcal{A}$ is equivalent to
replacing $D_{0}$ by $D$. For simplicity we shall write
\begin{equation}
D=\gamma ^{\mu }\left( \partial _{\mu }+A_{\mu }\right)
\end{equation}
where $A_{\mu }$ is traceless and $D$ acts only on the Hilbert space of
fermions, but not their conjugates. The bosonic action for this example was
worked out, up to terms with two derivatives, in \cite{ACAC}. It is given by
the expansion
\begin{equation}
Tr F\left( D^2\right) =\sum\limits_{n\geq 0}f_{n}a_{n}\left( D^2\right)
\end{equation}
where
\begin{eqnarray}
f_0 &=& \int_{0}^{\infty} uF(u) du ,\nonumber\\ f_2 &=&
\int_{0}^{\infty} F(u) du ,\nonumber\\ f_{2(n+2)} &=& (-1)^n
F^{(n)}(0),\qquad n\geq 0
\end{eqnarray} .
To obtain terms
with derivatives of order higher  than two one must consider the Seeley-de
Witt coefficient $a_{6}\left( D^{2}\right) $.
To find out these terms we first write
\begin{equation}
P=D^{2}=-(\partial ^{\mu }\partial _{\mu }+L^{\mu }\partial _{\mu }+B)
\end{equation}
where for simplicity we assumed a flat manifold $M$ with $g_{\mu
\nu }=\delta _{\mu \nu }$. For the space in (\ref{space}) we have
\begin{eqnarray}
L^{\mu } &=&2A^{\mu } \nonumber\\ B &=&\partial ^{\mu }A_{\mu
}+A^{\mu }A_{\mu }+\frac{1}{2}\gamma ^{\mu \nu }F_{\mu \nu } .
\end{eqnarray}
The result of Gilkey \cite{Gilkey} for $a_{6}\left( P\right) $ is:
\begin{eqnarray}
a_{6} &=&\frac{1}{360}Tr\,\left( 8F^{\mu \nu ;\rho }F_{\mu \nu ;\rho
}+2F_{\mu \nu }^{\quad ;\nu }F_{\quad ;\rho }^{\mu \rho }+12F_{\mu \nu
}F_{\quad \quad \rho }^{\mu \nu ;\rho }\right. \nonumber\\
&&\qquad \qquad -12F_{\mu \nu }F^{\nu \rho }F_{\rho }^{\mu }+6E_{;\mu \nu
}^{\quad \,\mu \nu }+60EE_{;\mu }^{\quad \mu } \nonumber\\
&&\qquad \qquad \left. +30E^{;\mu }E_{;\mu }+60E^{3}+30E\Omega _{\mu \nu
}\Omega ^{\mu \nu }\right)
\end{eqnarray}
where
\begin{eqnarray}
\omega ^{\mu } &=&\frac{1}{2}L^{\mu },\nonumber \\
\Omega _{\mu \nu } &=&\partial _{\mu }\omega _{\nu }-\partial _{\nu }\omega
_{\mu }+\left[ \omega _{\mu },\omega _{\nu }\right] ,\nonumber \\
E &=&B-\left( \partial ^{\mu }A_{\mu }+A^{\mu }A_{\mu }\right) .
\end{eqnarray}
In the simple example under consideration we have
\begin{eqnarray}
\omega ^{\mu } &=& A^{\mu } \nonumber\\
\Omega _{\mu \nu } &=& F_{\mu \nu } \nonumber\\
E &=&-\frac{1}{2}\gamma ^{\mu \nu }F_{\mu \nu } .
\end{eqnarray}

It is straightforward to work out the final form of $\int a_{6}\left(
P\right) d^{4}x\sqrt{g}$. After integrating by parts and ignoring boundary
terms, we obtain
\begin{equation}
\int\limits_{M}a_{6}=\frac{1}{360}\int\limits_{M}Tr\,\left( 11F^{\mu \nu
;\rho }F_{\mu \nu ;\rho }+2F_{\mu \nu }^{\quad ;\nu }F_{\quad ;\rho }^{\mu
\rho }+48F_{\mu \nu }F^{\nu \rho }F_{\,\,\rho }^{\mu }\right) \sqrt{g}d^{4}x .
\end{equation}
The last term vanishes in the abelian case $\left( N=1\right) $, as
this is proportional to
\begin{equation}
f^{abc}F_{\mu \nu }^{a}F^{\nu \rho b}F_{\rho }^{\,\,\mu c}
\end{equation}
where $f^{abc}$ are the structure constants of $SU(N)$.

\section{Supersymmetrisation of the bosonic terms}

To find the supersymmetric extension of the higher order bosonic
term in the heat-kernel expansion of the spectral action, the
easiest way is to use the method of superfields and write an action
in superspace. We shall use the notation of Bagger and Wess
\cite{Bag} .

Let $V\left( x^{\mu },\theta ^{\alpha
},\overline{\theta}^{\dot{\alpha} }\right) $ be a real vector
superfield. Introduce the covariant supersymmetry and gauge
operators
\begin{eqnarray}
\nabla _{\alpha } &=&e^{-V}D_{\alpha }e^{V} \nonumber\\
\overline{\nabla }_{\dot{\alpha}} &=&e^{-V}D_{\dot{\alpha}}e^{V} \nonumber\\
\nabla _{\mu } &=&e^{-V}\partial _{\mu }e^{V}
\end{eqnarray}
where $D_{\alpha }=\frac{\partial }{\partial \theta ^{\alpha }}+i(\sigma
^{\mu }\bar{\theta})_{\alpha }\partial _{\mu }$ and $\overline{D}_{\dot{%
\alpha}}=-\frac{\partial }{\partial \overline{\theta }_{\dot{\alpha}}}%
-i(\theta \sigma ^{\mu })_{\dot{\alpha}}\partial _{\mu }$ are the supersymmetry
covariant derivatives which satisfy
\begin{eqnarray}
\left\{ D_{\alpha },D_{\beta }\right\} &=&0=\left\{ \overline{D}_{\dot{\alpha%
}},\overline{D}_{\dot{\beta}}\right\} \nonumber\\
\left\{ D_{\alpha },\overline{D}_{\dot{\alpha}}\right\} &=&
-2i(\sigma ^{\mu})_{\alpha \dot{\alpha}}\partial _{\mu } .
\end{eqnarray}
Notice that $ \nabla_{\alpha }$ could be expressed in the alternative form
\begin{equation}
\nabla_{\alpha }=D_{\alpha }+\left( e^{-V}D_{\alpha}e^{V}\right)
\end{equation}
 where the expression within brackets is not an
operator. The operators $\nabla _{\alpha }$and $\overline{\nabla
}_{\dot{\alpha}}$ satisfy
\begin{eqnarray}
\left\{ \nabla _{\alpha },\nabla _{\beta }\right\} &=&0=\left\{ \overline{%
\nabla }_{\dot{\alpha}},\overline{\nabla }_{\dot{\beta}}\right\} \nonumber\\
\left\{ \nabla _{\alpha },\overline{\nabla }_{\dot{\alpha}}\right\}
&=&-2i(\sigma ^{\mu })_{\alpha \dot{\alpha}}\nabla _{\mu } .
\end{eqnarray}

The field $W_{\alpha }=\overline{D}^{2}\left( e^{-V}D_{\alpha }e^{V}\right) $
transforms covariantly under the transformation $e^{V}\rightarrow e^{-i%
\overline{\Lambda }}e^{V}e^{i\Lambda }$ where $\Lambda $ and $\overline{%
\Lambda }$ are chiral and antichiral fields ($\overline{D}_{\dot{\alpha%
}}\Lambda =0$):
\begin{equation}
W_{\alpha }\rightarrow e^{-i\Lambda }W_{\alpha }e^{i\Lambda }
\end{equation}
and therefore could be used in constructing a supersymmetric and gauge
invariant action.
In the Wess-Zumino gauge, the vector field $V$ takes the simple form
\begin{equation}
V=-\theta \sigma ^{\mu }\overline{\theta }A_{\mu }(x)+i\theta \theta
\overline{\theta }\overline{\lambda }-i\overline{\theta }\overline{\theta }%
\theta \lambda +\frac{1}{2}\theta \theta \overline{\theta }\overline{\theta }%
X .
\end{equation}
One can verify that $W_{\alpha }$ contains the gauge field strength
\cite{Bag}.

\begin{equation}
W_{\alpha }=-i\lambda _{\alpha }(y)+\theta _{\beta }\left( \delta _{\alpha
}^{\beta }X(y)-i\left( \sigma ^{\mu \nu }\right) _{\alpha }^{\beta }F_{\mu
\nu }(y)\right) +\theta \theta \left( \sigma ^{\mu }D_{\mu }\overline{%
\lambda }(y)\right) _{\alpha }
\end{equation}
where $y^{\mu }=x^{\mu }+i\theta \sigma ^{\mu }\overline{\theta }$.

A supersymmetric action that does not contain terms higher than two
derivatives is
\begin{equation}
\frac{1}{16}\int d^{2}\theta Tr\,\left( W^{\alpha }W_{\alpha }\right) =-%
\frac{1}{4}F_{\mu \nu }^{a}F^{\mu \nu a}-i\overline{\lambda }^{a}\gamma
^{\mu }D_{\mu }\lambda _{a}+\frac{1}{2}X^{a}X^{a}
\end{equation}
with an identical expression for the conjugate term.

To include terms with higher derivative, especially those  found in
the expression of $a_{6}\left( P\right) $ we consider the symmetric
and antisymmetric pieces of $\left\{ \nabla _{\alpha },W_{\beta
}\right\} $ as these transform covariantly
\begin{equation}
\left\{ \nabla _{\alpha },W_{\beta }\right\} \rightarrow e^{-i\Lambda
}\left\{ \nabla _{\alpha },W_{\beta }\right\} e^{i\Lambda } .
\end{equation}
The invariant actions are
\begin{eqnarray}
&&Tr\int d^{2}\theta d^{2}\overline{\theta }\left\{ \nabla ^{\alpha
},W_{\alpha }\right\} \left\{ \nabla ^{\beta },W_{\beta }\right\} +h.c
\label{symact}\\
&&Tr\int d^{2}\theta d^{2}\overline{\theta }\left( \left\{ \nabla ^{\alpha
},W^{\beta }\right\} +\left\{ \nabla ^{\beta },W^{\alpha }\right\} \right)
\left( \left\{ \nabla _{\alpha },W_{\beta }\right\} \right) +h.c \label{antsymm} .
\end{eqnarray}
The list of terms coming from the first action (\ref{symact}) are
\begin{eqnarray}
&&Tr\left( F_{\mu \nu ;}^{\quad \,\,\nu }F_{\quad ;\rho }^{\mu \rho }\right)
,\qquad Tr\left( D_{\mu }XD^{\mu }X\right) \nonumber\\
&&Tr\left( \overline{\lambda }\gamma ^{\mu }D_{\mu }D^{\nu }D_{\nu }\lambda
\right) ,\qquad Tr\left( \overline{\lambda }\gamma ^{\mu }\lambda \overline{%
\lambda }\gamma _{\mu }\lambda \right) \nonumber\\
&&Tr\left( D_{\mu }\overline{\lambda }\gamma ^{\mu }\gamma ^{\nu \rho
}\lambda F_{\nu \rho }\right) ,\qquad Tr\left( \overline{\lambda }\gamma
^{\mu }\lambda F_{\mu \nu ;}^{\quad \,\,\nu }\right) \nonumber\\
&&Tr\left( \overline{\lambda }\gamma ^{\mu }\lambda D_{\mu
}X\right) .
\end{eqnarray}
We notice the appearance of one term quartic in the fermions $Tr\left(
\overline{\lambda }\gamma ^{\mu }\lambda \ \overline{\lambda }\gamma _{\mu
}\lambda \right) $, which is of the current-current interaction
type. Similarly,  from the second action (\ref{antsymm}) we have
\begin{eqnarray}
&&Tr\left( F_{\mu \nu }F^{\nu \rho }F_{\rho }^{\quad \mu }\right) ,\qquad
Tr\left( F_{\mu \nu ;}^{\quad \,\,\nu }F_{\quad ;\rho }^{\mu \rho }\right)
\qquad Tr\left( F_{\mu \nu ;\rho }F^{\mu \nu ;\rho }\right) \nonumber\\
&&\qquad Tr\left( \overline{\lambda }\gamma ^{\mu \nu }\gamma
^{\rho }D_{\rho }\lambda F_{\mu \nu }\right) ,\qquad Tr\left(
\overline{\lambda }%
\gamma ^{\mu }\lambda \ \overline{\lambda }\gamma _{\mu }\lambda \right) .
\end{eqnarray}
Again the quartic interaction term appears. It is possible to fix
the coefficient between the two actions (\ref {symact}) and (\ref
{antsymm}) to cancel the quartic fermionic interaction. Without
evaluating the higer order terms, it is not clear whether this
phenomena of obtaining non-trivial fermionic interactions would
persist and whether one would be able to eliminate them.

We now compare these results with those obtained from the fermionic action
\begin{equation}
I_{f}=\left( \Psi \left| G\left( D\right) \right| \Psi \right)
\end{equation}
where $G\left( D\right) $ is an odd function of $D$
\begin{equation}
G\left( D\right) =\sum\limits_{n=0}^{\infty }g_{2n+1}D^{2n+1} .
\end{equation}
This implies that
\begin{equation}
I_{f}=\sum\limits_{n=0}^{\infty }g_{2n+1}\left( \Psi \left| D^{2n+1}\right|
\Psi \right) .
\end{equation}
Note that there is no loss of generality by assuming that $g\left( D\right) $
is odd as $\left( \Psi \left| D^{2n}\right| \Psi \right) $ is zero by
chirality. Evaluating $\left( \Psi \left| D^{3}\right| \Psi \right) $ we
obtain
\begin{equation}
\left( \Psi \left| D^{3}\right| \Psi \right) =\left( \Psi \left| -D^{\mu
}D_{\mu }\gamma ^{\rho }D_{\rho }\right| \Psi \right) +\frac{1}{2}\left(
\Psi \left| \gamma ^{\mu \nu }F_{\mu \nu }\gamma ^{\rho }D_{\rho }\right|
\Psi \right) .
\end{equation}
The field $X$ appearing in the supersymmetric theory could be
obtained by replacing $D$ by $D+X$. Therefore, one sees that the
simple expression for the fermionic action can recover all terms
appearing in the supersymmetric theory, to that order in
derivatives, except for the quartic fermionic terms which for
certain combinations can be made to vanish. It is therefore
natural, from the spectral action point of view where an action is
defined in terms of the Dirac operator, to postulate that the
fermionic action is given by $I_{f}$. It would be preferable to
have some symmetry principle that relates the functions $F$ and $G$
appearing in the bosonic and fermionic terms to each other. In the
spectral action of the superstring \cite{AHC} the same expression
generates both the bosonic and the fermionic actions.
Unfortunately, in two-dimensions one has to work with world-sheet
fermions and not space-time fermions which are connected in a
non-transperent way,  and this makes it difficult to obtain such a
relation.

\section{Conclusions}

In this work we have studied two questions. The first is the issue
of chiral fermions in the spectral action, and whether these could
be introduced consistently, as the eigenvalue problem is not well
defined. The trace has to be restricted to the set of
eigenfunctions for the operator $D_{-}D_{+}$. This makes the trace
non-invariant under chiral rotations, resulting in  gauge and
gravitational anomalies. We obtained a set of conditions for
cancellation of anomalies which coincided with the known
conditions. The reasons for appearance of anomalies are, however,
differnt. In the path integral formulation, chiral anomalies arise
because of the lack of invariance of the chiral fermions measure
under chiral rotations. The second question we addressed is the
form of the fermionic spectral action, and whether one can find a
simple form for it. We studied supersymmetrisation of terms
containing derivatives of orders higher than two, appearing in the
bosonic effective spectral action. We found that there are, in
general, quartic fermionic terms which could not be written in a
way that depends only on the operator $D$ as they are of the
current-current type. These terms could be arranged to cancel.
 It is not possible to write the quartic fermionic term in a local
way as a function of the operator $D$. One can simply adopt the
point of view, that the fermionic action must be given by a simple
form, quadratic in fermions, and dependent on some odd function of
$D$. We therefore conjecture that the spectral action is given by
\begin{equation}
I=Tr\left( F\left( D^{2}/m^{2}\right) \right) +\left( \Psi \left| G\left(
D\right) \right| \Psi \right)
\end{equation}
It is very likely that a relation exists between the functions $F$
and $G$, but without a guiding symmetry principle, such a relation
is not easy to find.
\section{Acknowledgments}

I would like to thank Alain Connes for correspondence.


\begin{thebibliography}{9}
\bibitem{Connes} A. Connes, {\it Publ. Math. IHES} {\bf 62} (1983) 44.

\bibitem{Cobook} A. Connes, {\it Noncommutative Geometry}, Academic Press,
New York, 1994.

\bibitem{CoLo} A. Connes and J. Lott, {\it Nucl. Phys. Proc.} {\bf B18} (1990)
295.

\bibitem{Kastler} D. Kastler, {\it Rev. Math. Phys. } {\bf 5} (1993) 477.

\bibitem{CFF} A. H. Chamseddine, G. Felder and J. Fr\"ohlich, {\it Nucl. Phys. }
{\bf B395} (1992) 672; {\it Phys. Lett.} {\bf 296B} (1992) 109.

\bibitem{Co95} A. Connes, {\it J. Math. Phys. } {\bf 36} (1995) 6255.

\bibitem{ACAC}  A. H. Chamseddine and A. Connes, {\it Phys. Rev. Lett.} {\bf 77}
(1996) 4868; {\it Commun. Math. Phys. } {\bf 186} (1997) 731.

\bibitem{Fuji}  K. Fujikawa, {\it Phys. Rev. Lett}. {\bf 44} (1980)
1733; {\it Phys. Rev. } {\bf D21 } (1980) 2848; {\bf D29} (1984)
285.

\bibitem{Bala}  A. P. Balachandran, G. Marmo, V. P. Nair and C. G. Trahern,
{\it Phys. Rev.} {\bf D25} (1983) 2713.

\bibitem{Alv}  L. Alvarez-Gaum\'{e} and P. Ginsparg, {\it Nucl. Phys.}
{\bf B243} (1984) 449.

\bibitem{Gilkey}  P. B. Gilkey \textit{Invariance Theory, The Heat Equation
and the Atiyah-Singer Index Theorem} Publish or Perish Inc. (1984),
Washington, Delaware (USA).

\bibitem{A-W}  L. Alvarez-Gaum\`{e} and E. Witten, {\it Nucl. Phys.}
{\bf B234} (1984) 269.

\bibitem{Bag}  J. Bagger and J. Wess, \textit{Supersymmetry and Supergravity }
Princeton University Press, Second edition (1991).

\bibitem{AHC}  A. H. Chamseddine, {\it Phys. Rev.} {\bf D56} (1997) 3555.

\end{thebibliography}
\end{document}